\documentclass[aps,prd,onecolumn,groupedaddress,showpacs,nofootinbib,amssymb
]{revtex4}
\usepackage[dvips]{graphicx}
\usepackage{amssymb}
\usepackage{amsmath}
\usepackage{graphicx}
\usepackage{amsfonts}
\usepackage{bm}

\begin{document}

\title{Dark Energy Oscillations in Mimetic $F(R)$ Gravity}
\author{
S.~D.~Odintsov,$^{1,2}$\,\thanks{odintsov@ieec.uab.es}
V.~K.~Oikonomou,$^{3,4}$\,\thanks{v.k.oikonomou1979@gmail.com}}
\affiliation{ $^{1)}$Institute of Space Sciences (IEEC-CSIC), Cr.Can Magrans s/n,
08193 Barcelona, Spain\\
$^{2)}$ ICREA, Passeig Luis Companys, 23,
08010 Barcelona, Spain\\
$^{3)}$ Tomsk State Pedagogical University, 634061 Tomsk, Russia\\
$^{4)}$ Laboratory for Theoretical Cosmology, Tomsk State University of Control Systems
and Radioelectronics (TUSUR), 634050 Tomsk, Russia\\
}

\begin{abstract}
In this paper we address the problem of dark energy oscillations in the context of mimetic $F(R)$ gravity with potential. The issue of dark energy oscillations can be a problem in some models of ordinary $F(R)$ gravity and a remedy that can make the oscillations milder is to introduce additional modifications in the functional form of the $F(R)$ gravity. As we demonstrate the power-law modifications are not necessary in the mimetic $F(R)$ case, and by appropriately choosing the mimetic potential and the Lagrange multiplier, it is possible to make the oscillations almost to vanish at the end of the matter domination era and during the late-time acceleration era. We examine the behavior of the dark energy equation of state parameter and of the total effective equation of state parameter as functions of the redshift and we compare the resulting picture with the ordinary $F(R)$ gravity case. As we also show, the present day values of the dark energy equation of state parameter and of the total effective equation of state parameter are in better agreement with the observational data, in comparison to the ordinary $F(R)$ gravity case. Finally, we study the evolution of the growth factor as a function of the redshift for all the mimetic models we shall use.

\end{abstract}


\maketitle



\def\pp{{\, \mid \hskip -1.5mm =}}
\def\cL{\mathcal{L}}
\def\be{\begin{equation}}
\def\ee{\end{equation}}
\def\bea{\begin{eqnarray}}
\def\eea{\end{eqnarray}}
\def\tr{\mathrm{tr}\, }
\def\nn{\nonumber \\}
\def\e{\mathrm{e}}

\section{Introduction}

The mimetic gravity idea was introduced some time ago by Mukhanov and Chamseddine \cite{mukhanov1}, and ever since it has been thoroughly studied in the literature \cite{mukhanov2,Golovnev1,Golovnev2,Golovnev3,Golovnev4,Golovnev5,Golovnev6,Golovnev7,NO2,mimetic1,mimetic2,mimetic3,mimetic4,mimetic5}. In the context of mimetic gravity, the conformal degree of freedom of the background metric becomes a matter component of the full theory, so this could play the role of cold dark matter. Later on in Ref. \cite{NO2}, the mimetic gravity theory was studied in the context of $F(R)$ gravity, and the formalism of Lagrange multipliers \cite{CMO1,CMO3} was used. The presence of the Lagrange multiplier and of the potential offered the possibility for realizing various cosmologies, see for example Refs. \cite{mimetic4,mimetic8}, for some recent studies.

In this paper we shall use the mimetic $F(R)$ gravity framework in order to investigate the behavior of dark energy oscillations in the presence of the potential and Lagrange multiplier. In the standard $F(R)$ gravity approach, there is a big issue coming from dark energy oscillations \cite{staro} near the end of the matter domination era and during the late-time acceleration era \cite{exp0,exp1,exp2,exp3a,exp4}. This oscillations issue persists in some models even at present time, so this is a rather disturbing issue. In the standard approach, in order to make the dark energy oscillations milder, it was compelling to modify the $F(R)$ gravity by adding by hand certain power-law modifications. As we will show, in the context of mimetic $F(R)$ gravity, it is not necessary to add any power-law modifications and also by appropriately choosing the mimetic potential and Lagrange multiplier, this can almost make the oscillations to disappear or to have very small amplitude near and during the late-time era.

In this paper we shall be interested in mimetic $F(R)$ gravity models, in which we will specify the mimetic potential, the Lagrange multiplier and the $F(R)$ gravity. With regards to the $F(R)$ gravity, we shall choose an exponential model which was introduced in \cite{importantpapers11}. This exponential model passes all the local tests \cite{reviews1,reviews2,reviews3,reviews4,reviews5,rev6} and some of its variants unify early
and late-time acceleration. Also it mimics the $\Lambda$CDM model, so it has appealing properties. In addition, the motivation for using exponential $F(R)$ gravity models comes from the fact that recent studies indicate that exponential models resemble to a great extent the $\Lambda$CDM model \cite{chen}. Particularly the data coming from baryon acoustic oscillations in the clustering of galaxies and also studies of the shift parameters corresponding to the cosmic microwave background radiation indicate that there is no possible way to distinguish the resulting cosmological evolutions caused by exponential $F(R)$ models and the $\Lambda$CDM model \cite{chen}. This justifies to a great extent why we decided to use an exponential $F(R)$ gravity model. Our aim is to use special variables which measure the deviation of the mimetic $F(R)$ gravity model from the $\Lambda$CDM model, and we shall study the dark energy equation of state parameter $\Omega_{DE}(z)$ and of the total equation of state parameter $\omega_{eff}(z)$ as functions of the redshift $z$. For this calculation we shall use specific forms of the mimetic potential and Lagrange multiplier and also we shall use the exponential $F(R)$ gravity without the power-law modifications. The result is compared to the exponential  with power-law term ordinary $F(R)$ gravity case, and as we demonstrate the oscillations in the mimetic case are damped and these have smaller amplitude in comparison to the ordinary $F(R)$ gravity case. Finally, we compute the growth factor of matter perturbations for both the mimetic and non-mimetic $F(R)$ gravity. For the calculation we shall assume that the Universe is filled with radiation only in the mimetic case and with non-relativistic matter and radiation in the ordinary $F(R)$ case. The resulting picture shows that the evolution of the growth factor as a function of the redshift is different for the various mimetic models when compared to the ordinary $F(R)$ model and also the mimetic models have differences between them. Finally, we briefly discuss various alternative cosmological scenarios with one of these being related to the cosmographic approach.

We need to note that the results of our analysis strongly indicate that the phantom line is crossed even in the context of mimetic $F(R)$ theory. In the literature, a lot of effort has been made towards the complete understanding of the $w=-1$ line crossing, in the context of dark energy physics, see for example Ref. \cite{ref1} for a comprehensive review on this issue. This issue captivates the interest of cosmologists, due to an existing proof of the no-go theorem on this issue, see for example Ref. \cite{ref2}, which states that in the standard Einstein-Hilbert gravity, the cosmological perturbations may become ill behaved if the dark energy crosses the phantom line $w=-1$. Now this issue may be overcome in the context of mimetic $F(R)$ gravity, and in principle a suitable choice of the potential may indeed succeed that, however in this paper we will not address this physically ``deep'' issue, since a concise treatment of the cosmological perturbations in the context of mimetic $F(R)$ gravity is needed, which could be treated as a special form of an $F(R,\phi)$ gravity. However this issue is significantly interesting and we will address it in a future work, especially the impact of the phantom divide crossing on the cosmological perturbations of mimetic $F(R)$ gravity. In addition to the perturbations issue, in principle, quantum field theory becomes inconsistent with phantom cosmology due to a catastrophic instability that occurs in the theory. Also phantom cosmology maybe expected to occur in the future, but it is hard to believe it happened in the past (nevertheless, there are proposals on phantom inflation). In addition, oscillatory behavior in the total effective equation of state may also quickly bring instabilities, so this is something that needs to be avoided. As we shall demonstrate mimetic gravity succeeds in making the oscillations almost to vanish, but the phantom divide crossing occurs, so a deeper analysis is needed to a fundamental level to see if mimetic gravity makes the phantom dived crossing to disappear from the theory. This would be interesting, since only a complete theory of quantum gravity makes this possible.

This paper is organized as follows: In section II we present in brief the mimetic $F(R)$ gravity formalism and also the essentials of the exponential $F(R)$ gravity model. Then we present the general formalism for studying the dark energy oscillations and we perform a numerical analysis for various mimetic $F(R)$ models. Also we study in brief the evolution of the growth factor as a function of the redshift and we discuss some alternative scenarios related to the cosmographic principle. Finally, the conclusions follow in the end of the paper.

\section{Dark Energy AND Equation of State Oscillations and Mimetic $F(R)$ Gravity Models}

\subsection{Mimetic $F(R)$ Gravity Essentials}

Here we shall present the mimetic $F(R)$ gravity framework with Lagrange multiplier and potential. For more details on these issues see \cite{NO2,mimetic4,mimetic5}. In the context of mimetic $F(R)$ gravity, the hidden internal conformal degrees of freedom of the metric are being used \cite{mukhanov1}, and therefore the physical metric $g_{\mu \nu}$ can be written in terms of an auxiliary scalar field $\phi$ and in terms of an auxiliary metric $\hat{g}_{\mu \nu}$, in the following way
\begin{equation}\label{metrpar}
g_{\mu \nu}=-\hat{g}^{\rho \sigma}\partial_{\rho}\phi \partial_{\sigma}\phi
\hat{g}_{\mu \nu}\, .
\end{equation}
The most important assumption in the context of mimetic theories is that the gravitational action should be varied with respect to the auxiliary metric. From Eq. (\ref{metrpar}) it easily follows that,
\begin{equation}\label{impl1}
g^{\mu \nu}(\hat{g}_{\mu \nu},\phi)\partial_{\mu}\phi\partial_{\nu}\phi=-1\,
.
\end{equation}
Also as it can easily be checked, the parametrization (\ref{metrpar}) is invariant under the Weyl transformation $\hat{g}_{\mu \nu}=e^{\sigma (x)}g_{\mu \nu}$, and effectively the auxiliary metric $\hat{g}_{\mu
\nu}$ does not appear in the final action. We assume that the physical metric is a Friedmann-Robertson-Walker (FRW) metric with line element,
\begin{equation}\label{frw}
ds^2 = - dt^2 + a(t)^2 \sum_{i=1,2,3}
\left(dx^i\right)^2\, ,
\end{equation}
with $a(t)$ being as usual the scale factor. The gravitational action of the mimetic $F(R)$ gravity with mimetic potential $V(\phi )$ and Lagrange multiplier $\lambda (\phi )$ is \cite{NO2},
\begin{equation}\label{actionmimeticfraction}
S=\int \mathrm{d}x^4\sqrt{-g}\left ( F\left(R(g_{\mu \nu})\right
)-V(\phi)+\lambda \left(g^{\mu \nu}\partial_{\mu}\phi\partial_{\nu}\phi
+1\right)\right )+L_{matt}\, ,
\end{equation}
where $L_{matt}$ is the Lagrangian of the matter fields present, which we shall assume that are perfect fluids. Also we assume that the auxiliary scalar field depends only on the cosmic time. By varying the action (\ref{actionmimeticfraction}) with respect to the metric tensor $g_{\mu \nu}$, we obtain the following equations,
\begin{align}\label{aeden}
& \frac{1}{2}g_{\mu \nu}F(R)-R_{\mu
\nu}F'(R)+\nabla_{\mu}\nabla_{\nu}F'(R)-g_{\mu \nu}\square F'(R)\\ \notag &
+\frac{1}{2}g_{\mu \nu}\left (-V(\phi)+\lambda \left( g^{\rho
\sigma}\partial_{\rho}\phi\partial_{\sigma}\phi+1\right) \right )-\lambda
\partial_{\mu}\phi \partial_{\nu}\phi +\frac{1}{2}T_{\mu \nu}=0 \, ,
\end{align}
where $T_{\mu \nu}$ is the energy momentum tensor corresponding to the matter fluids present. Moreover, by varying the action with respect to the auxiliary scalar field $\phi$, we obtain,
\begin{equation}\label{scalvar}
-2\nabla^{\mu} (\lambda \partial_{\mu}\phi)-V'(\phi)=0\, ,
\end{equation}
where the ``prime'' denotes differentiation with respect to the auxiliary scalar $\phi$. Upon variation of the action (\ref{actionmimeticfraction}) with respect to the Lagrange multiplier $\lambda$, we obtain,
\begin{equation}\label{lambdavar}
g^{\rho \sigma}\partial_{\rho}\phi\partial_{\sigma}\phi=-1\, ,
\end{equation}
and as it can be seen, the above equation is identical to Eq. (\ref{impl1}). Hence the mimetic constraint (\ref{impl1}) is satisfied also for the mimetic theory with Lagrange multiplier and mimetic potential. Note that the Lagrange multiplier actually introduces the mimetic constraint, so this is why in the present case we varied the action (\ref{actionmimeticfraction}) with respect to the physical metric and not with respect to the auxiliary metric. For the metric (\ref{frw}), and under the assumption that the scalar field depends on the cosmic time, the equations of motion (\ref{aeden}), (\ref{scalvar})
and (\ref{lambdavar}), can be written in the following way,
\begin{equation}\label{enm1}
-F(R)+6(\dot{H}+2H^2)F'(R)-6H\frac{\mathrm{d}F'(R)}{\mathrm{d}t}-\lambda
(\dot{\phi}^2+1)+V(\phi)+\rho_{matter}=0\, ,
\end{equation}
\begin{equation}\label{enm2}
F(R)-2(\dot{H}+3H^2)+2\frac{\mathrm{d}^2F'(R)}{\mathrm{d}t^2}+4H\frac{\mathrm{d}F'(R)}{\mathrm{d}t}-\lambda (\dot{\phi}^2-1)-V(\phi)+p_{matter}=0\, ,
\end{equation}
\begin{equation}\label{enm3}
2\frac{\mathrm{d}(\lambda \dot{\phi})}{\mathrm{d}t}+6H\lambda
\dot{\phi}-V'(\phi)=0\, ,
\end{equation}
\begin{equation}\label{enm4}
\dot{\phi}^2-1=0\, ,
\end{equation}
where the ``dot'' denotes differentiation with respect to the cosmic time $t$, and the prime in Eqs. (\ref{enm1}) and (\ref{enm2}) denotes differentiation with respect to the Ricci scalar $R$, while in Eq. (\ref{enm3}) it denotes differentiation with respect to the auxiliary scalar\footnote{Note that in the absence of the mimetic potential, the FRW equations are identical to the ordinary $F(R)$ gravity plus the mimetic term $C_{\phi}/a^{3}$, as it can be seen in Ref. \cite{NO2}.}. Also $\rho_{matter}$ and $p_{matter}$ are the energy density and the effective pressure of the matter fluids present. From Eq. (\ref{enm3}) it follows that the auxiliary scalar is identified with the cosmic time $t$,  a feature also present in the Einstein-Hilbert mimetic gravity, as it can easily be seen by the mimetic constraint (\ref{impl1}). Hence, by using this identification, Eq. (\ref{enm2}) can be written in the following way,
\begin{equation}\label{sone}
F(R)-2(\dot{H}+3H^2)+2\frac{\mathrm{d}^2F'(R)}{\mathrm{d}t^2}+4H\frac{\mathrm{d}F'(R)}{\mathrm{d}t}-V(t)+p_{matter}=0\, .
\end{equation}
Effectively, the mimetic potential $V(t)$ is expressed in terms of the $F(R)$ gravity, the Hubble rate and the effective pressure as follows,
\begin{equation}\label{scalarpot}
V(\phi=t)=2\frac{\mathrm{d}^2F'(R)}{\mathrm{d}t^2}+4H\frac{\mathrm{d}F'(R)}{
\mathrm{d}t}+F(R)-2(\dot{H}+3H^2)+p_{matter}\, .
\end{equation}
Correspondingly, the Lagrange multiplier $\lambda (t)$ is equal to,
\begin{equation}\label{lagrange}
\lambda (t)=-3 H \frac{\mathrm{d}F'(R)}{\mathrm{d}t}+3
(\dot{H}+H^2)-\frac{1}{2}F(R)+\rho_{matter}\, .
\end{equation}
Given the Hubble rate $H(t)$, it is possible to realize any arbitrary cosmic evolution. In addition, given the $F(R)$ gravity and the Hubble rate, by appropriately choosing the Lagrange multiplier and the mimetic potential, it is possible to realize a viable cosmological evolution. Our purpose in this paper is to investigate how the dark energy oscillations behave in the presence of the mimetic potential and the Lagrange multiplier. As we shall see, the mimetic potential and Lagrange multiplier introduce many new appealing features in the theory under study. Before we proceed to this, in the next section we shall briefly present the essential features of the $F(R)$ we shall use.

\subsection{Exponential with power-law term $F(R)$ Gravity and Cosmological Viability}

In the literature there exist various viable $F(R)$ gravity models, which have to satisfy a number of constraints in order these can be considered viable \cite{reviews1,reviews2,reviews3,reviews4,reviews5,rev6}. For the purposes of this paper we shall use an exponential $F(R)$ gravity model which has very appealing features, and it was introduced in Ref. \cite{importantpapers11}. Particularly, it successfully passes all the local astrophysical constraints, it provides a theoretical framework for the unified description of the late and early-time acceleration, and it mimics the $\Lambda$CDM model at large curvature values. For detailed studies of exponential models of $F(R)$ gravity see \cite{importantpapers11,exp0,exp1,exp2,exp3a,exp4,exponentialmodels4,exponentialmodels5,exponentialmodels7}. The model of Ref. \cite{importantpapers11} which we will study in this paper, has the following functional form,
\begin{equation}\label{expmodnocurv}
F(R)=R-2\Lambda \left ( 1-e^{\frac{R}{b\Lambda}} \right )\, ,
\end{equation}
where $\Lambda$ is the present time cosmological constant and $b$ is a positive free parameter which is assumed to be $\mathcal{O}(1)$.

An issue that occurs in $F(R)$ gravity theories is the existence of large dark energy oscillations during the matter domination era and later on. The higher derivatives of the Hubble rate strongly diverge as a consequence of the dark energy oscillations during the matter domination era, as was discussed in detail in Refs. \cite{exp0,exp1,exp4}. Particularly, the high frequency dark energy oscillations occur for large $z$, where $z$ is the cosmological redshift which is related to the scale factor as $a=1/(z+1)$. In addition, the derivatives of the dark energy effective energy density, which we denote as $\rho_{DE}$, also take large values, and in effect this procedure affects the dark energy equation of state parameter $\omega_{DE}$. Moreover, the dark energy oscillations have higher frequency in the cases that the $F(R)$ gravity mimics the $\Lambda$CDM model, which occurs when $F''(R)\simeq 0$. This issue also occurs for the model (\ref{expmodnocurv}), so a consistent way to remedy the high curvature regime of the model (\ref{expmodnocurv}) was proposed in Ref. \cite{exp1}, were power-law modifications were included in the $F(R)$ gravity (\ref{expmodnocurv}). Particularly, the exponential  with power-law term $F(R)$ gravity has the following form,
\begin{equation}\label{expmodnocurvcorr}
F(R)=R-2\Lambda \left ( 1-e^{\frac{R}{b\Lambda}} \right )-\tilde{\gamma}\Lambda \left (\frac{R}{3\tilde{m}^2} \right )^{1/3}
\end{equation}
where $\Lambda=7.93\tilde{m}^2$ and $\tilde{\gamma}=1/1000$, see Ref. \cite{exp1} for details. In effect the dark energy oscillations are stabilized and the frequency of the oscillations becomes nearly constant \cite{exp1}. Notice that with the addition of the power-law modifications, the viability of the model is retained, while the dark energy oscillations problem is amended too. For example the flat space solution is still the Minkowski spacetime, and the power-law modifications vanish in the de-Sitter epoch, when $\tilde{\gamma}$ satisfies $\tilde{\gamma}\ll (\tilde{m}^2/\Lambda)^{1/3} $, which is satisfied by the choice of $\tilde{\gamma}$ we made earlier.

As we already mentioned in the introduction, in order to stabilize the dark energy oscillations in the non-mimetic $F(R)$ gravity case, the power-law modifications are necessary, at least for the class of models under study. However, as we demonstrate in the next sections, the power-law modifications are not needed in the case of the mimetic $F(R)$ gravity since by appropriately choosing the mimetic potential and the Lagrange multiplier, the high frequency oscillations of the dark energy disappear or become damped, at least for the specific class of $F(R)$ models under study. Thus, the mimetic potential and the Lagrange multiplier play the role of the power-law modifications and more importantly, these appear from the beginning in the theory without the need to introduce them by hand. In the next section we demonstrate in detail how this can be done in the context of mimetic $F(R)$ gravity.

\subsection{Dark Energy Oscillations Formalism in Mimetic $F(R)$ Gravity}

In this section we shall study the dark energy oscillations of a Universe filled with radiation in the context of mimetic $F(R)$ gravity, and we compare these oscillations to the exponential  with power-law term $F(R)$ gravity case oscillations. As we already mentioned, in the standard $F(R)$ gravity approach, the dark energy oscillations have extremely large frequencies near the end of the matter domination era. This is not an appealing feature since these oscillations would affect even the present time era, so in order to amend this issue, in the standard $F(R)$ gravity approach, power-law modifications are needed and in effect the oscillations are damped at the late stages of our Universe evolution. As we demonstrate in this section, there is no need for power-law modifications in the mimetic $F(R)$ case, since the mimetic potential and the Lagrange multiplier actually amend the dark energy oscillations. As we will show, the choice of the potential and Lagrange multiplier plays an important role. Also in some cases the oscillations have almost zero amplitude and the values of the dark energy density and of the effective equation of state parameter at present time are very close to observational values.

For the purposes of our study, we shall express all the physical quantities as functions of the redshift $z=1/a-1$ and also we shall introduce new variables which are more appropriate for the study of the dark energy oscillations. The resulting differential equations will determine the behavior and evolution of the dark energy oscillations, and by numerically solving these we will study the evolution of the dark energy density $\Omega_{DE}(z)$ and of the effective equation of state parameter $\omega_{eff}(z)$, as functions of the redshift $z$. For the numerical analysis the focus will be for redshifts $z\leq 10$, since we are interested for the last stage of the matter domination era, which started at $z\sim 3000$ and finished around $z\sim 3$. After this era, the late-time acceleration era started which continues until today at $z\sim 0$. There is a specific motivation for us to express all the physical quantities as functions of the redshift, and we specify for values $0\leq z\leq 10$, since the near future observational data will come from Gamma Ray Bursts or Type Ia supernovae or other standard candles, corresponding to redshifts with $z\geq 6$.

Let us proceed to construct the master equation which governs the dark energy oscillations evolution. We rewrite the FRW equation (\ref{enm1}) as follows,
\begin{align}\label{eq:flrw}
& 3F'H^2=\frac{\rho_{matt}}{2}+\frac{V(t)-2\lambda(t)}{2}+\frac{1}{2}(F'R-F)-3H\dot{F'}\, ,
\end{align}
where $\rho_{matt}$ is the total mass energy density of the perfect fluids present. For the mimetic $F(R)$ gravity case, we shall assume that only radiation is present, although even for the vacuum case, the resulting picture is qualitatively the same. So for the mimetic case, the total energy density is,
\begin{equation}\label{totalmattenergdensmf1}
\rho_{matt}=\rho_{r}^{(0)}a^{-4}\, ,
\end{equation}
while for the non-mimetic standard $F(R)$ gravity case, the energy density will be assumed to be as follows,
 \begin{equation}\label{totalmattenergdensmf}
\rho_{matt}=\rho_m^{(0)}a^{-3}+\rho_{r}^{(0)}a^{-4}\, .
\end{equation}

Rewriting the first equation in Eq. (\ref{eq:flrw}) we obtain,
\begin{equation}\label{eq:modifiedeinsteineqns2}
  H^2-(F'-1)\left (H\frac{\mathrm{d}H}{\mathrm{d}\ln{a}}+H^2 \right )+\frac{1}{6}(F-R)+H^2F''\frac{\mathrm{d}R}{\mathrm{d}\ln{a}}-\frac{V(t)-2\lambda(t)}{3}=\frac{\rho_{matt}}{3},
\end{equation}
with $R$ being the Ricci scalar which can be written as follows,
\begin{equation}\label{eq:ricciscal2}
  R=12H^2+6H\frac{\mathrm{d}H}{\mathrm{d}\ln{a}}.
\end{equation}
For notational simplicity we introduce the following function,
\begin{equation}\label{potlagra}
  \mathcal{Q}(a(z))=V(a(z))-2\lambda (a(z))\, ,
\end{equation}
so if this is specified, this will determine how the oscillations will evolve, as we now demonstrate. Notice that if the function $\mathcal{Q}(z)$ is specified, then by using Eq. (\ref{enm3}), one easily finds the Lagrange multiplier, given the Hubble rate and via $\mathcal{Q}(z)$, the potential can be determined.

The dark energy oscillations can be consistently described by using the following variables,
\begin{subequations}
  \begin{align}
    y_H&\equiv\frac{\rho_{DE}}{\rho_m^{(0)}}=\frac{H^2}{\tilde{m}^2}-\mathcal{Q}(a)-\chi a^{-4}\, ,\label{eq:yH}\\
    y_R&\equiv\frac{R}{\tilde{m}^2}-\frac{\mathrm{d}\mathcal{Q}(a)}{\mathrm{d}\ln{a}}\, ,\label{eq:yR}
  \end{align}
\end{subequations}
with $\rho_{DE}$ being the total energy density of dark energy, and also $\rho_m^{(0)}$ and $\rho_r^{(0)}=\chi\rho_m^{(0)}$ are the present values of the mass-energy densities corresponding to non interacting matter and radiation respectively. In addition, the variable $\chi$ is the ratio $\chi=\rho_r^{(0)}/\rho_m^{(0)}$. The new variable that can describe the dark energy oscillations is called the scaled dark energy $y_H(z)$. To proceed, we divide Eqs. (\ref{eq:modifiedeinsteineqns2}) by $\tilde{m}^2$, and by using,
\begin{equation}
  \frac{H^2}{\tilde{m}^2}-\frac{\rho_M}{2\tilde{m}^2}=\frac{H^2}{\tilde{m}^2}-\mathcal{Q}(a)-\chi a^{-4}=y_H
\end{equation}
from (\ref{eq:yH}), we can easily solve the resulting expression with respect to the term $\frac{1}{\tilde{m}^2}\frac{dR}{d\ln{a}}$. Then, upon differentiation of  Eq. (\ref{eq:yR}), with respect to the expression $\ln{a}$ we get,
\begin{equation}\label{eq:dyR}
  \frac{\mathrm{d}y_R}{\mathrm{d}\ln{a}}=-\frac{\mathrm{d}^2\mathcal{Q}(a)}{\mathrm{d}\ln{a}^2}+\left[-y_H+(F'-1)(\frac{H}{\tilde{m}^2}\frac{\mathrm{d}H}{\mathrm{d}\ln{a}}
    +\frac{H^2}{\tilde{m}^2})-\frac{1}{6\tilde{m}^2}(F-R)\right]\frac{1}{H^2F''}
\end{equation}
Then by using Eq. (\ref{eq:yH}), differentiating Eq. (\ref{eq:yH}) with respect to $\ln{a}$, and finally using Eq. (\ref{eq:dyR}) we acquire,
\begin{eqnarray}\label{eq:dyR2}
  \frac{\mathrm{d}y_R}{\mathrm{d}\ln{a}}&=&-\frac{\mathrm{d}^2\mathcal{Q}(a)}{\mathrm{d}\ln{a}^2}+
    \left[-y_H+(F'-1)\left(\frac{1}{2}\frac{\mathrm{d}y_H}{\mathrm{d}\ln{a}}+\frac{1}{2}\frac{\mathrm{d}\mathcal{Q}(a)}{\mathrm{d}\ln{a}}
    +y_H+\mathcal{Q}(a)-\chi a^{-4}\right)\right.\nonumber\\
    &&\left.-\frac{1}{6\tilde{m}^2}(F-R)\right]\frac{1}{\tilde{m}^2F''(y_H+\mathcal{Q}(a)+\chi a^{-4})}.
\end{eqnarray}
Differentiating Eq. (\ref{eq:dyR2}) with respect to $\ln{a}$ we get,
\begin{equation}\label{eq:dyH}
\frac{\mathrm{d}y_H}{\mathrm{d}\ln{a}}=\frac{2H}{\tilde{m}^2}\frac{\mathrm{d}H}{\mathrm{d}\ln{a}}-\frac{\mathrm{d}\mathcal{Q}(a)}{\mathrm{d}\ln{a}}+4\chi a^{-4}.
\end{equation}
By combining Eqs. (\ref{eq:ricciscal2}) and (\ref{eq:yH}), in effect Eq. (\ref{eq:dyH}) becomes,
\begin{eqnarray}\label{eq:dyH2}
  \frac{\mathrm{d}y_H}{\mathrm{d}\ln{a}}&=&\frac{R}{3\tilde{m}^2}-\frac{\mathrm{d}\mathcal{Q}(a)}{\mathrm{d}\ln{a}}-4y_H-4\mathcal{Q}(a)\nonumber\\
    &=&\frac{y_R}{3}-4y_H-\frac{2\mathrm{d}\mathcal{Q}(a)}{3\mathrm{d}\ln{a}}-4\mathcal{Q}(a)
\end{eqnarray}
and moreover we rewrite the Ricci scalar as follows,
\begin{equation}
  R=3\tilde{m}^2\left(4y_H+4\mathcal{Q}(a)+\frac{\mathrm{d}y_H}{\mathrm{d}\ln{a}}+\frac{\mathrm{d}\mathcal{Q}(a)}{\mathrm{d}\ln{a}}\right)
\end{equation}
Differentiating Eq. (\ref{eq:dyH2}) with respect to $\ln{a}$, and also by making use of Eq. (\ref{eq:dyR2}), we acquire,
\begin{eqnarray}\label{eq:FRform}
  &&\frac{\mathrm{d}^2y_H}{\mathrm{d}\ln{a}^2}+\left(4+\frac{1-F'}{6\tilde{m}^2F''(y_H+\mathcal{Q}(a)+
    \chi a^{-4})}\right)\frac{\mathrm{d}y_H}{\mathrm{d}\ln{a}}
    +\left(\frac{2-F'}{3\tilde{m}^2F''(y_H+\mathcal{Q}(a)+\chi a^{-4})}\right)y_H
  \nonumber\\
  &&+\left(\frac{\mathrm{d}^2\mathcal{Q}(a)}{\mathrm{d}\ln{a}^2}+4\frac{\mathrm{d}\mathcal{Q}(a)}{\mathrm{d}\ln{a}}
    +\frac{(1-F')\left(3\frac{\mathrm{d}\mathcal{Q}(a)}{\mathrm{d}\ln{a}}+6\mathcal{Q}(a)-6\chi a^{-4}\right)+\frac{F-R}{\tilde{m}^2}}
      {18\tilde{m}^2F'' (y_H+\mathcal{Q}(a)+\chi a^{-4})}\right)=0
\end{eqnarray}
Having Eq. (\ref{eq:FRform}) we can easily express all the physical quantities as functions of the redshift $z$ by using the following two rules,
\begin{subequations}
  \begin{align}
    \frac{\mathrm{d}}{\mathrm{d}\ln{a}}&=-(z+1)\frac{\mathrm{d}}{\mathrm{d}z},\\
    \frac{\mathrm{d}^2}{\mathrm{d}\ln{a}^2}&=(z+1)\frac{\mathrm{d}}{\mathrm{d}z}+(z+1)^2\frac{\mathrm{d}^2}{\mathrm{d}z^2}\, ,
  \end{align}
\end{subequations}
so the dark energy oscillations in a Universe filled with radiation and governed my a mimetic $F(R)$ gravity theory, are described by the following master equation,
\begin{eqnarray}\label{mastereqn}
  &&\frac{\mathrm{d}^2y_H}{\mathrm{d}z^2}+
    \frac{1}{(z+1)}\left(-3-\frac{F'(R)}{6\tilde{m}^2F''(R)(y_H+\mathcal{Q}(z)+\chi(z+1)^4)}\right)\frac{\mathrm{d}y_H}{\mathrm{d}z}
  \nonumber\\
  &&+\frac{1}{(z+1)^2}\frac{1-F'(R)}{3\tilde{m}^2F''(R)(y_H+\mathcal{Q}(z)+\chi(z+1)^4)}y_H+\left(\frac{
    \mathrm{d}^2\mathcal{Q}(z)}{\mathrm{d}z^2}-\frac{3}{(z+1)}\frac{\mathrm{d}\mathcal{Q}(z)}{\mathrm{d}z}\right.
  \nonumber\\
  &&\left.-\frac{1}{(z+1)}\frac{F'(R)\left(-(z+1)\frac{\mathrm{d}\mathcal{Q}(z)}{
    \mathrm{d}z}+2\mathcal{Q}(z)-2\chi(z+1)^4\right)+\frac{F}{3\tilde{m}^2}}{6\tilde{m}^2F''(R)(y_H-\mathcal{Q}(z)+\chi(z+1)^4)}\right)=0\, .
\end{eqnarray}
We will solve numerically this differential equation by specifying the model $\mathcal{Q}(z)$ and by using the following initial conditions,
\begin{equation}\label{initialcond}
y_H(z)\mid_{z=z_{f}}=\frac{\Lambda }{3\tilde{m}^2}\left(1+\frac{z_{f}+1}{1000}\right),{\,}{\,}{\,}y_{H}'(z)\mid_{z=z_{f}}=\frac{\Lambda }{3\tilde{m}^2}\frac{1}{1000}
\end{equation}
with $z_{f}=10$ and $\Omega_M=0.279$. For the mimetic $F(R)$ gravity case, we shall assume that the $F(R)$ gravity will be that of Eq. (\ref{expmodnocurv}), and we shall compare the dark energy oscillations with the non-mimetic $F(R)$ gravity case, in which case the $F(R)$ gravity is the exponential  with power-law term appearing in Eq. (\ref{expmodnocurvcorr}). As we shall see, the mimetic $F(R)$ gravity leads to damped dark energy oscillations without the need of power-law modifications, and the observational features of the models are somewhat appealing.


\subsubsection{Numerical Analysis for Various $\mathcal{Q}(z)$ Models}

In this section we perform a numerical analysis of the differential equation (\ref{mastereqn}) which determines the evolution of the dark energy oscillations in the Universe, in the context of mimetic $F(R)$ gravity, using the numerical values and initial conditions we specified in the previous section.
 \begin{figure}[h]
\centering
\includegraphics[width=15pc]{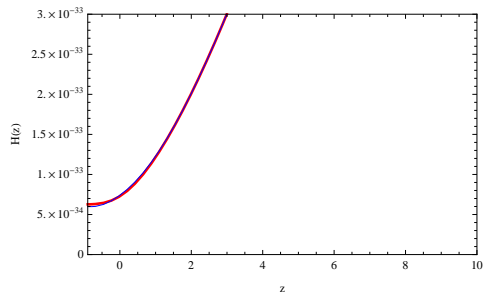}
\caption{Comparison of the Hubble parameter $H(z)$ over z for the model $\mathcal{Q}(z)=\sqrt{2 z+5}$} \label{plot1}
\end{figure}
We shall specify the mimetic potential and Lagrange multiplier by providing the exact form of the function $\mathcal{Q}(z)$ given in Eq. (\ref{potlagra}). For the mimetic $F(R)$ gravity case, the $F(R)$ gravity will be the one appearing in Eq. (\ref{expmodnocurv}), which has no power-law modifications and we compare the resulting picture with the ordinary $F(R)$ gravity in which case the $F(R)$ gravity is given by (\ref{expmodnocurvcorr}) and it has power-law modifications. Our analysis will involve the evolution of the Hubble rate $H(z)$, which in terms of $y_H(z)$ is equal to,
\begin{equation}\label{hubblepar}
H(z)=\sqrt{\tilde{m}^2(y_H(z)+\mathcal{Q}(z)+\chi (z+1)^{4})}\, .
\end{equation}
In addition we study the dark energy density $\omega_{DE}=P_{DE}/\rho_{DE}$, which is equal to,
\begin{equation}\label{deeqnstateprm}
\omega_{DE}(z)=-1+\frac{1}{3}(z+1)\frac{1}{y_H(z)}\frac{\mathrm{d}y_H(z)}{\mathrm{d}z}\, ,
\end{equation}
and finally the total equation of state parameter $\omega_{eff}(z)$, which is equal to,
 \begin{equation}\label{effeqnofstateform}
\omega_{eff}(z)=-1+\frac{2(z+1)}{3H(z)}\frac{\mathrm{d}H(z)}{\mathrm{d}z}
\end{equation}
We start off the analysis with the model,
\begin{equation}\label{model1}
\mathcal{Q}(z)=\sqrt{2z+5}\, ,
\end{equation}
\begin{figure}[h]
\centering
\includegraphics[width=15pc]{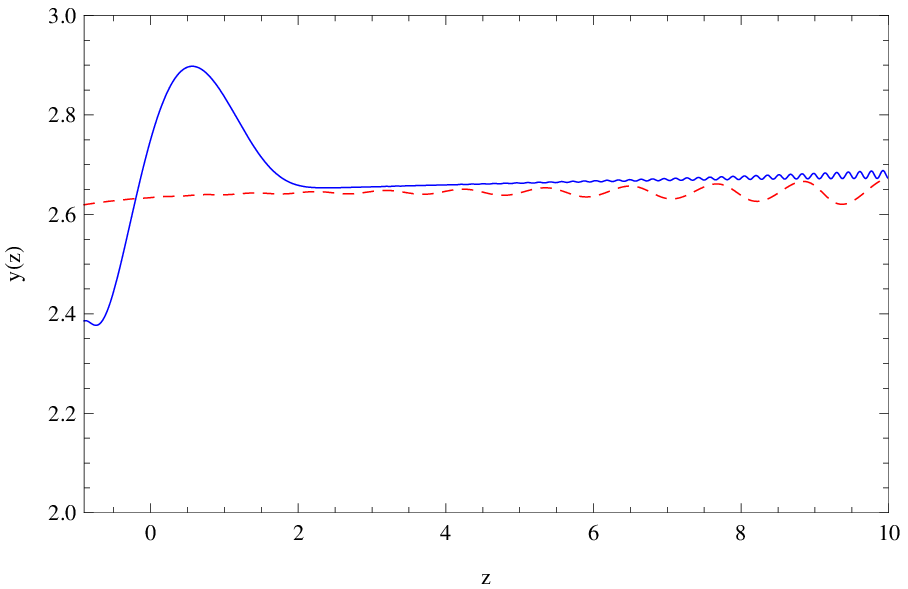}
\includegraphics[width=15pc]{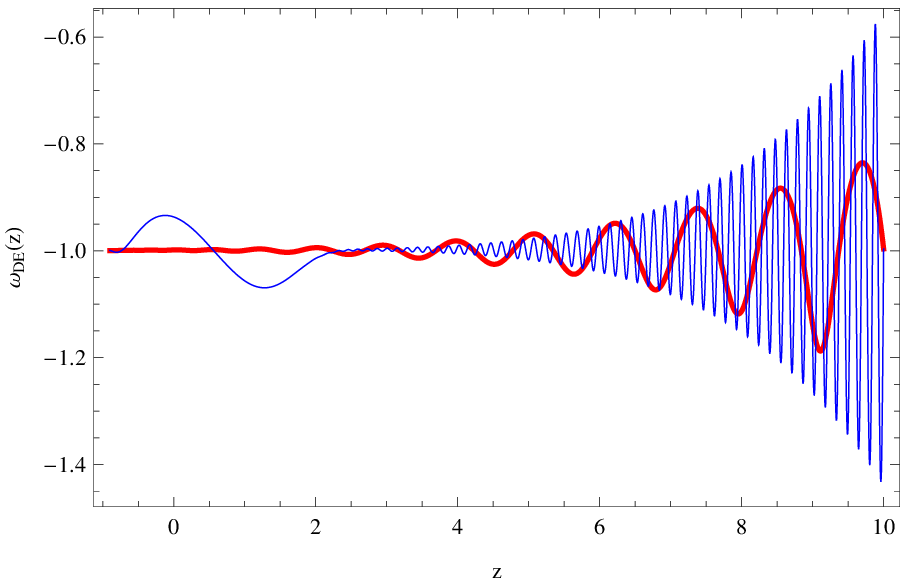}
\includegraphics[width=15pc]{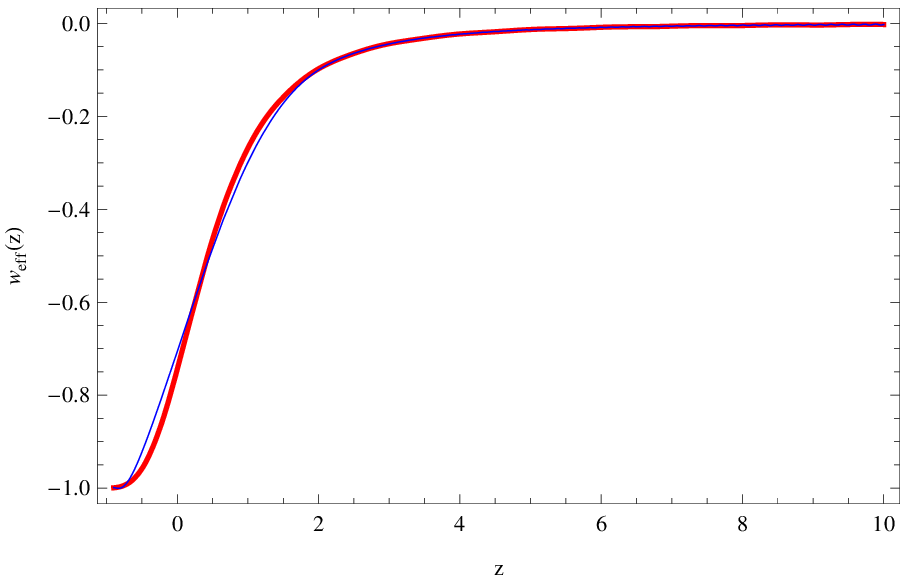}
\caption{Comparison of the scaled dark energy density $y_H(z)=\frac{\rho_{DE}}{\rho_m^{(0)}}$ over z (left plot), of the dark energy equation of state parameter $\omega_{DE}(z)$ over z (right plot) and of the effective equation of state parameter $\omega_{eff}(z)$ over z  (bottom plot)  for the model $\mathcal{Q}(z)=\sqrt{2 z+5}$. The red curves correspond to the mimetic $F(R)$ case, while the blue curves correspond to the exponential  with power-law term ordinary $F(R)$ model.} \label{plot2}
\end{figure}
and in Fig. \ref{plot1} we plot the evolution of the Hubble rate as a function of the redshift $z$, for the mimetic $F(R)$ model (\ref{model1}), (red curve) and for the ordinary $F(R)$ case (blue curve). As it can be seen in Fig. \ref{plot1} the two evolutions are practically indistinguishable, so the two theoretical frameworks yield the same cosmological evolution at the level of the Hubble rate. The differences can be found when the dark energy equation of state $\Omega_{DE}(z)$ and the scaled dark energy density $y_H(z)$ are studied, since higher derivatives of the Hubble rates are involved.
\begin{table*}
\small
\caption{\label{tablei} The values of the dark energy equation of state parameter $\omega_{DE}(z)$ and of the effective equation of state parameter $\omega_{eff}$, at present time for the model $\mathcal{Q}(z)=\sqrt{2 z+5}$ and for the ordinary $F(R)$ exponential  with power-law term model.}
\begin{tabular}{@{}crrrrrrrrrrr@{}}
\tableline
\tableline
\tableline
Model &  $\omega_{DE}(0)$ &  $\omega_{eff}(0)$
\\\tableline
Ordinary $F(R)$ Model &  $0.733292$ & $-0.936629$
 \\\tableline
$\mathcal{Q}(z)=\sqrt{2 z+5}$  &  $0.724737$ & $-0.998281$
\\\tableline
Observational Data &  $0.721\pm 0.015$ & $-0.972\pm 0.06$
\\\tableline
\tableline
 \end{tabular}
\end{table*}
In Fig. \ref{plot2} we present the comparison of the mimetic scaled dark energy density $y_H(z)=\frac{\rho_{DE}}{\rho_m^{(0)}}$ over z (left plot, red curve), of the mimetic dark energy equation of state parameter $\Omega_{DE}(z)$ over z (right plot, red curve) and of the mimetic effective equation of state parameter $\omega_{eff}(z)$ over z  (bottom plot, red curve), with the corresponding ordinary $F(R)$ gravity case (blue curves).
 \begin{figure}[h]
\centering
\includegraphics[width=15pc]{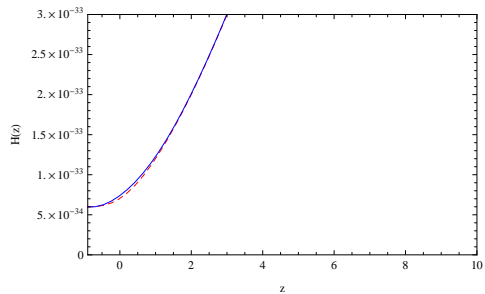}
\caption{Comparison of the Hubble parameter $H(z)$ over z for the model $\mathcal{Q}(z)=\frac{2 z+5}{z+100}$} \label{plot3}
\end{figure}
As it can be seen, the mimetic model (\ref{model1}) has very appealing properties since the scaled dark energy density has some oscillating behavior until approximately $z\sim 5$, at which point the oscillations are damped to a great extent. In addition, the dark energy equation of state parameter $\omega_{DE}(z)$ has oscillating behavior until $z\sim 3$ and after that there is no significant oscillating behavior, in contrast to the ordinary $F(R)$ case.
\begin{figure}[h]
\centering
\includegraphics[width=15pc]{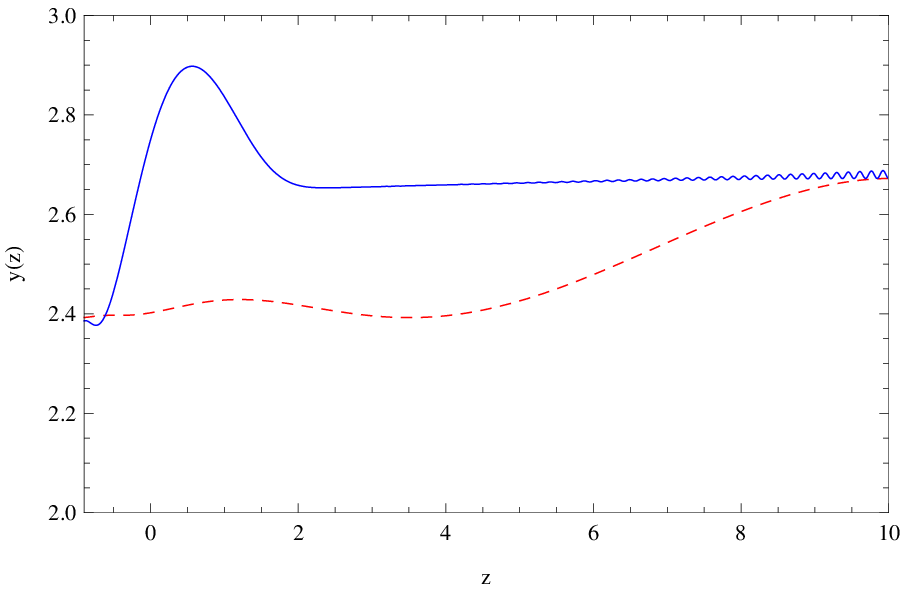}
\includegraphics[width=15pc]{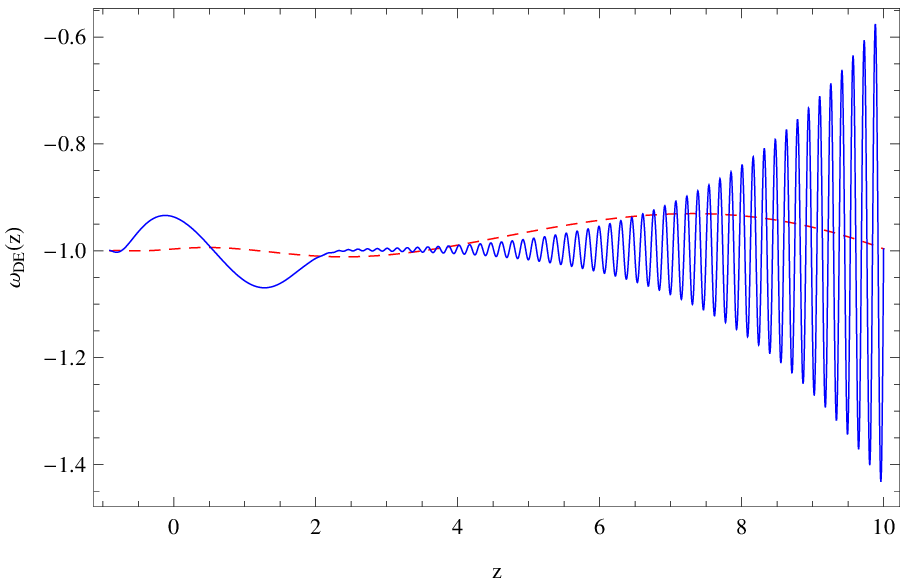}
\includegraphics[width=15pc]{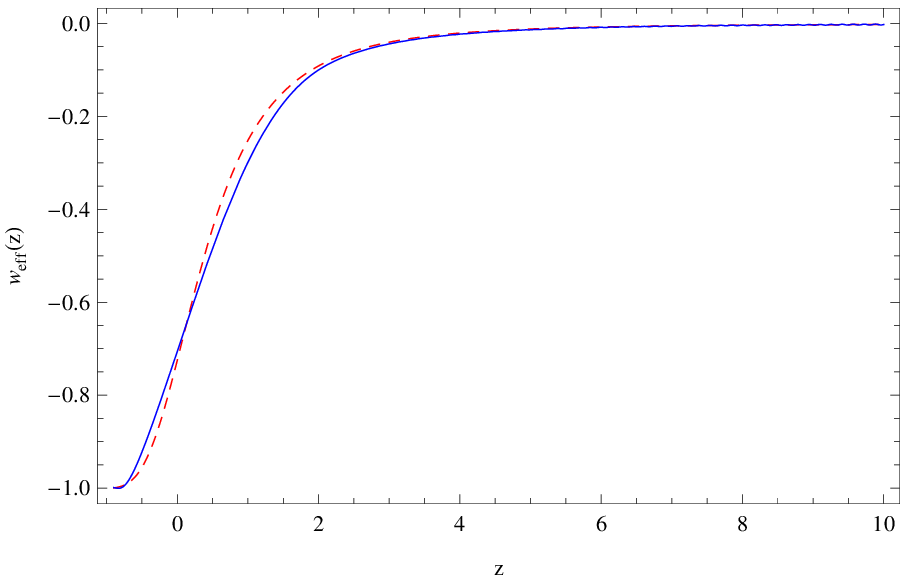}
\caption{Comparison of the scaled dark energy density $y_H(z)=\frac{\rho_{DE}}{\rho_m^{(0)}}$ over z (left plot), of the dark energy equation of state parameter $\omega_{DE}(z)$ over z (right plot) and of the effective equation of state parameter $\omega_{eff}(z)$ over z  (bottom plot)  for the model $\mathcal{Q}(z)=\frac{2 z+5}{z+100}$. The red curves correspond to the mimetic $F(R)$ case, while the blue curve corresponds to the exponential  with power-law term ordinary $F(R)$ model.} \label{plot4}
\end{figure}
Moreover, the oscillating behavior is milder in comparison to the exponential  with power-law term $F(R)$ gravity case. With regards to the effective equation of state parameter, as it can be seen in the bottom plot of Fig. \ref{plot2}, the two curves are very similar. An important feature of both cases is that the total equation of state never seems to cross the phantom divide line, or at least it is very close to it. Later on in this section we shall discuss again this issue. Also, in the mimetic case, the oscillations of the dark energy equation of state parameter around $-1$ have almost zero amplitude, which is an intriguing feature.
 \begin{table*}
\small
\caption{\label{tableii} The values of the dark energy equation of state parameter $\omega_{DE}(z)$ and of the effective equation of state parameter $\omega_{eff}$, at present time for the model $\mathcal{Q}(z)=\frac{2 z+5}{z+100}$ and for the ordinary $F(R)$ exponential  with power-law term model.}
\begin{tabular}{@{}crrrrrrrrrrr@{}}
\tableline
\tableline
\tableline
Model &  $\omega_{DE}(0)$ &  $\omega_{eff}(0)$
\\\tableline
Ordinary $F(R)$ Model &  $0.733292$ & $-0.936629$
 \\\tableline
$\mathcal{Q}(z)=\frac{2 z+5}{z+100}$  &  $0.706$ & $-0.996532$
\\\tableline
Observational Data &  $0.721\pm 0.015$ & $-0.972\pm 0.06$
\\\tableline
\tableline
 \end{tabular}
\end{table*}
Moreover in order to make contact with the observational data at present time, in Table \ref{tablei} we present the values of the dark energy density $\Omega_{DE}(z)$ and of the effective equation of state parameter $\omega_{eff}(z)$, at $z=0$, for the mimetic $F(R)$ and the ordinary $F(R)$ case.
\begin{table*}[h]
\small
\caption{\label{tableiii} The values of the dark energy equation of state parameter $\omega_{DE}(z)$ and of the effective equation of state parameter $\omega_{eff}$, at present time for the model $\mathcal{Q}(z)=\tanh (z+1)$ and for the ordinary $F(R)$ exponential  with power-law term model.}
\begin{tabular}{@{}crrrrrrrrrrr@{}}
\tableline
\tableline
\tableline
Model &  $\omega_{DE}(0)$ &  $\omega_{eff}(0)$
\\
\tableline
Ordinary $F(R)$ Model &  $0.733292$ & $-0.936629$
 \\
 \tableline
$\mathcal{Q}(z)=\tanh (z+1)$  &  $0.724737$ & $-0.998281$
\\
\tableline
Observational Data &  $0.721\pm 0.015$ & $-0.972\pm 0.06$
\\
\tableline
\tableline
 \end{tabular}
\end{table*}
As it can be seen, the mimetic case yields intriguingly appealing results since the predicted values are very close to the observational data, especially the dark energy equation of state parameter $\Omega_{DE}(0)$.

\begin{figure}[h]
\centering
\includegraphics[width=15pc]{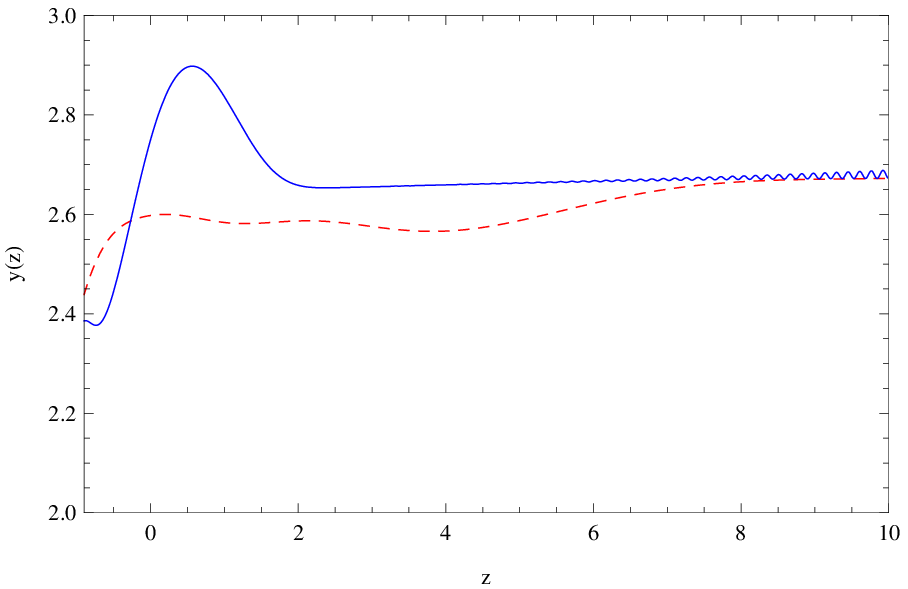}
\includegraphics[width=15pc]{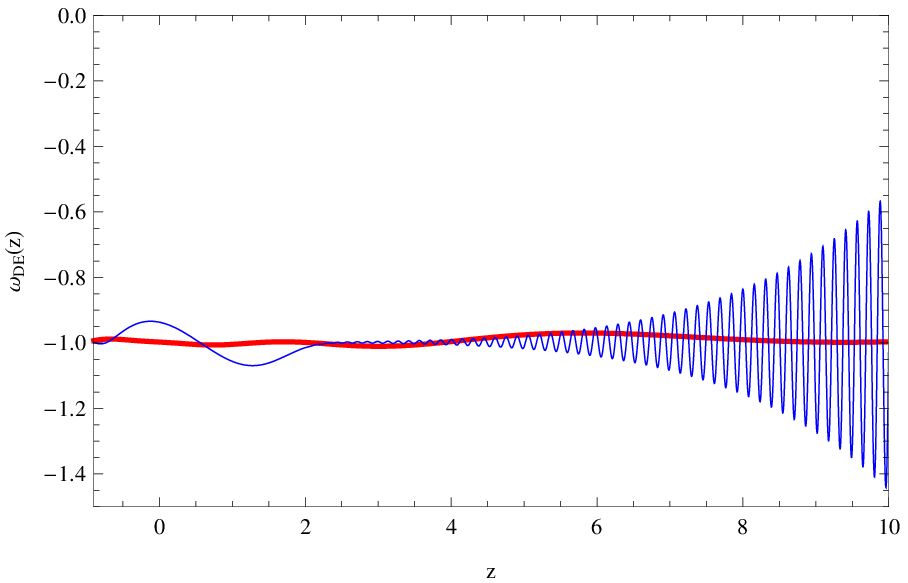}
\includegraphics[width=15pc]{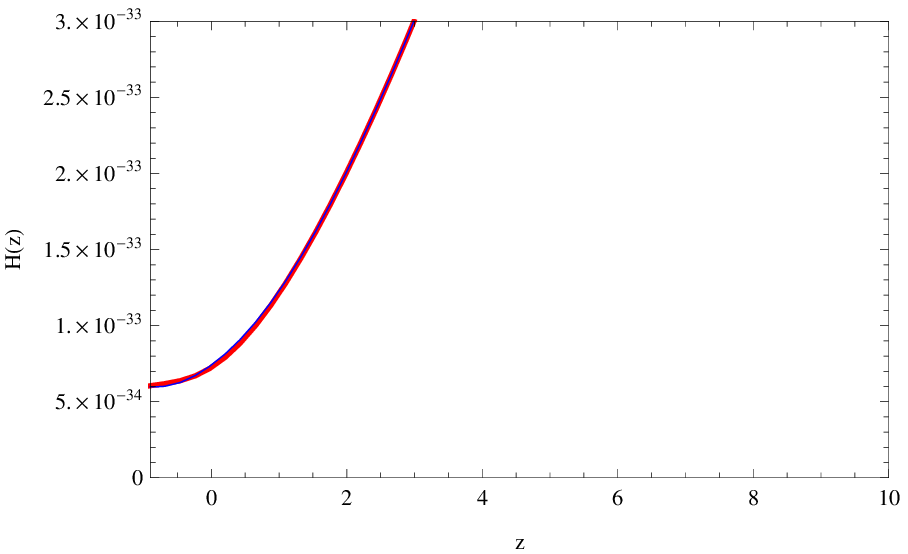}
\includegraphics[width=15pc]{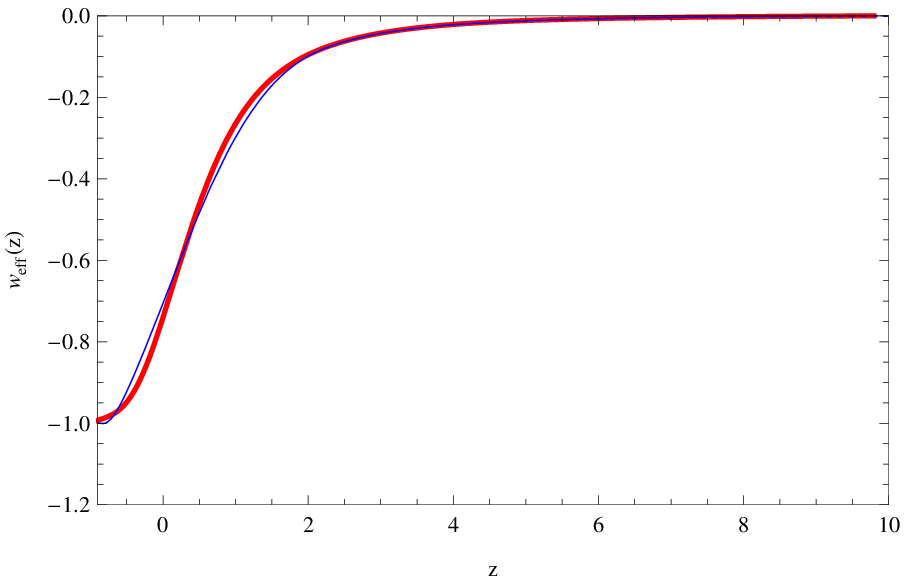}
\caption{Comparison of the scaled dark energy density $y_H(z)=\frac{\rho_{DE}}{\rho_m^{(0)}}$ over z (left plot), of the dark energy equation of state parameter $\omega_{DE}(z)$ over z (right plot) and of the effective equation of state parameter $\omega_{eff}(z)$ over z  (bottom plot right)  for the model $\mathcal{Q}(z)=\tanh (z+1)$. Also the Hubble rate can be found in the bottom left plot.} \label{plot5}
\end{figure}

The same qualitative picture can be obtained for other mimetic $F(R)$ models, for example the model,
\begin{equation}\label{model2}
\mathcal{Q}(z)=\frac{2 z+5}{z+100}\, ,
\end{equation}
yields also damped oscillating behavior for $\Omega_{DE}$ and also for the scaled dark energy density $y_H(z)$. The corresponding behaviors of the Hubble rate $H(z)$ and of $y_H(z)$, $\Omega_{DE}(z)$ and $\omega_{eff}(z)$ as functions of the redshift can be found in Fig. \ref{plot3} and Fig. \ref{plot4}.
Also in Table \ref{tableii} we present the predicted values of the aforementioned physical quantities. As it can be seen, the qualitative behavior is more or less the same in comparison to the model (\ref{model1}), however the observational values are not in such a good agreement as were the ones of the model (\ref{model1}). This feature is however model dependent, for example for the model,
\begin{equation}\label{model3}
\mathcal{Q}(z)=\tanh (z+1)\, ,
\end{equation}
the observational values of $\Omega_{DE}$ and $\omega_{eff}$ are in good agreement with the predicted one corresponding to the mimetic case, as it can be seen in Table \ref{tableiii}. The corresponding Hubble rate behavior and also the rest of the physical quantities can be found in Fig. \ref{plot5}.
In conclusion the mimetic $F(R)$ gravity models have the following appealing features:
\begin{itemize}
    \item The Hubble rate of the mimetic model is indistinguishable from the exponential  with power-law term ordinary $F(R)$ model.
    \item There is no need for the introduction of power-law modifications in the $F(R)$ function, since the oscillations are damped by choosing the potential and Lagrange multiplier.
    \item The dark energy equation of state oscillations have almost zero amplitude after $z\sim 3$, so after the end of the matter domination there is no issue of dark energy oscillations.
    \item The predicted values of the dark energy equation of state $\Omega_{DE}$ and the of total equation of state parameter $\omega_{eff}$ are very close to the observational values, in comparison to the ordinary $F(R)$ model.

\end{itemize}
Therefore a mimetic $F(R)$ gravity is distinguished from the ordinary $F(R)$ only if someone evaluates the predicted scaled dark energy density and the dark energy equation of state parameter. The rest of the physical quantities are more or less very similar in the two cases. Hence the question is whether there exists any other physical quantity which can clearly indicate sound differences between the mimetic and non-mimetic case, but also to indicate the differences between the various mimetic models. In the next section we study one such physical observable related to the matter density perturbations, namely the growth factor.

Before proceeding to the growth factor issue, it is worth discussing whether the phantom divide is crossed in the context of mimetic $F(R)$ gravity. For a detailed study on the phantom divide issue see \cite{phantomdiv}. We shall be interested for redshifts $0<z<3$, which correspond to the end of the matter domination era ($z\sim 3$) until today $z\sim 0$. In Table \ref{compphantom} we present the values of the total equation of state $\omega_{eff}$ for various redshifts in the interval $0<z<3$. As it can be seen, both the mimetic $F(R)$ models and the ordinary $F(R)$ model, cross the phantom divide line.
\begin{table*}[h]
\small
\caption{\label{compphantom} The values of the total effective equation of state parameter $\omega_{eff}$, for the mimetic $F(R)$ models and for the ordinary $F(R)$ model, for redshifts in the interval $0<z<3$.}
\begin{tabular}{@{}crrrrrrrrrrr@{}}
\tableline
\tableline
\tableline
Model &  $z=3$ &  $z=2.5$ & $z=2$ & $z=1.5$ & $z=1$ & $z=0.5$
\\\tableline
Ordinary $F(R)$ Model, $\omega_{eff}(z)$ &  $-0.998875$ & $-1.00003$ & $-1.00478$ & $-1.02439$ & $-1.02813$ & $-0.993667$
 \\\tableline
$\mathcal{Q}(z)=\tanh (z+1)$, $\omega_{eff}(z)$  &  $-1.00255$ & $-1.0017$ & $-0.999444$ & $-0.998923$ & $-1.00212$ & $-1.00303$
\\\tableline
$\mathcal{Q}(z)=\frac{2 z+5}{z+100}$, $\omega_{eff}(z)$  &  $-0.988715$ & $-0.990299$ &  $-0.993416$ & $-0.99778$&  $-1.00208$&  $-1.00335$
\\\tableline
$\mathcal{Q}(z)=\sqrt{2z+1}$, $\omega_{eff}(z)$  &  $-0.998685$ & $-1.00073$ & $-0.999282$ & $-0.999515$ & $-1.00006$ & $-0.999164$
\\\tableline
\tableline
 \end{tabular}
\end{table*}
Note that the phantom divide crossing usually leads to finite time future singularities, Type I possibly. So it would be interesting to investigate the effect of the mimetic potential on this type of singular evolutions, or even to add some extra imperfect fluid in the theory in order to avoid the future singularity. These tasks however exceed the purpose of this paper so we will address these in detail in a future work.

\subsection{Study of the Growth Index and Comparison with non-Mimetic Theories}

The cosmological perturbations are always a strong criterion that can help drastically to distinguish the various theoretical proposals for the Universe evolution. This is owing to the fact that the cosmological perturbations differentiate the predicted evolutions of each model from the background evolution, and this provides information on how this differentiation is achieved \cite{matsumoto}. The focus in this paper is on matter density perturbations, and we shall use the sub-horizon approximation. Practically, in the context of the sub-horizon approximation, the comoving wavelength $\lambda =a/k$ along with the spacelike hypersurface that describes the evolution, are much shorter than the corresponding Hubble horizon $R_H=1/(aH)$, or in a quantitative way this implies that $\frac{k^2}{a^2}\gg H^2$ with $k$ and $a$ being the comoving wavenumber an the scale factor respectively. For the mimetic $F(R)$ gravity case, the perfect fluid matter content is that of Eq. (\ref{totalmattenergdensmf1}), so the matter density perturbations are quantified by the parameter $\delta =\frac{\delta \varepsilon_m}{\varepsilon_m}$, where $\epsilon_m\sim a^{-4}$. Notice that this does not mean that the evolution is always radiation dominated, due to the presence of the mimetic potential and Lagrange multiplier.  The matter energy perturbation parameter $\delta$ satisfies the following differential equation,
\begin{equation}\label{matterperturb}
\ddot{\delta}+2H\dot{\delta}-4\pi G_{eff}(a,k)\varepsilon_m\delta =0
\end{equation}
with $G_{eff}(a,k)$ being the effective gravitational constant of the $F(R)$ theory given by \cite{exp1,dobado},
\begin{equation}\label{geff}
G_{eff}(a,k)=\frac{G}{8\pi F'(R)}\Big{[}1+\frac{\frac{k^2}{a^2}\frac{F''(R)}{F'(R)}}{1+3\frac{k^2}{a^2}\frac{F''(R)}{F'(R)}} \Big{]}
\end{equation}
with $G$ being the Newton constant. Notice that in the mimetic $F(R)$ case with potential, the effective gravitational constant is also given by  Eq. (\ref{geff}), but it is worth providing a brief proof of this. The mimetic $F(R)$ theory of gravity is a theory of the form $F(R,\phi, X)$, with $X$ being the kinetic term of the scalar field $\phi$, that is $X=-g^{\mu \nu}\partial_{\mu}\phi\partial_{\nu}\phi $. For this kind of theories the effective gravitational constant in the subhorizon approximation is \cite{tsuji1},
\begin{equation}\label{correctgef}
G_{eff}(a,k)=\frac{G}{8\pi F_R}\frac{F_{X}+4\left(F_X \frac{k^2}{a^2}\frac{F_{RR}}{F_R}+\frac{F_{R\phi}}{F_R}\right)}{F_{X}+3\left(F_X \frac{k^2}{a^2}\frac{F_{RR}}{F_R}+\frac{F_{R\phi}}{F_R}\right)}\, ,
\end{equation}
where $F_R$ is the the derivative of $F(R,\phi, X)$ with respect to $R$, $F_X$ with respect to $X$ and so on. Eventually in the mimetic $F(R)$ case, since $\partial_{\phi} F(R)=0$, the expression (\ref{correctgef}) becomes identical to (\ref{geff}). The physical quantity we shall study in this section is the growth factor $f_g(z)=\frac{\mathrm{d}\ln \delta}{\mathrm{d}\ln a}$, so expressing the differential equation (\ref{matterperturb}) in terms of the growth factor and the redshift, we obtain,
\begin{equation}\label{presenceofcoll}
\frac{\mathrm{d}f_g(z)}{\mathrm{d}z}+\Big{(}\frac{1+z}{H(z)}\frac{\mathrm{d}H(z)}{\mathrm{d}z}-2-f_g(z)\Big{)}\frac{f_g(z)}{1+z}+\frac{4\pi}{G}\frac{G_{eff}(a(z),k)}{(z+1)H^2(z)}\varepsilon_m=0
\end{equation}
The contribution of the mimetic potential and Lagrange multiplier comes from the Hubble rate and also notice that in the mimetic case the $F(R)$ gravity is different from the ordinary $F(R)$ gravity, since there are no power-law modifications.
\begin{figure}[h]
\centering
\includegraphics[width=15pc]{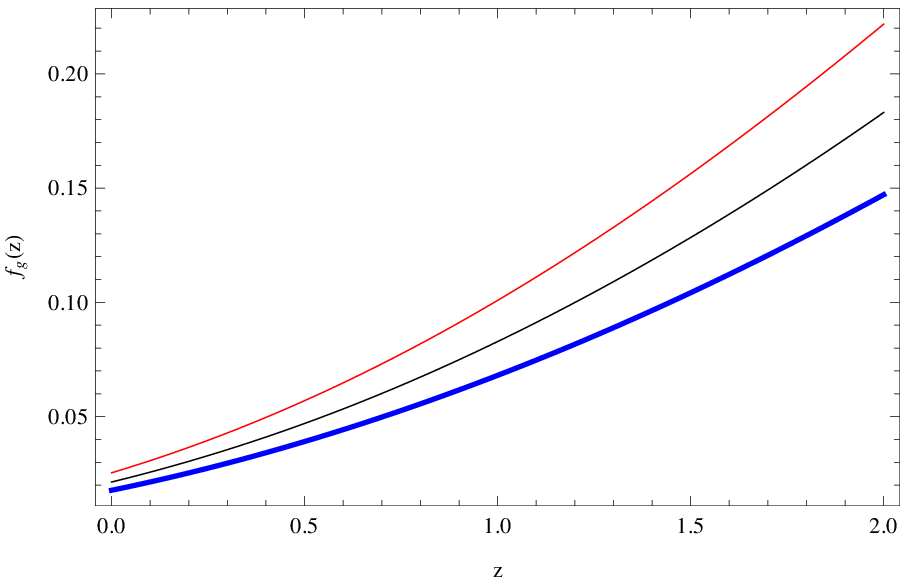}
\includegraphics[width=15pc]{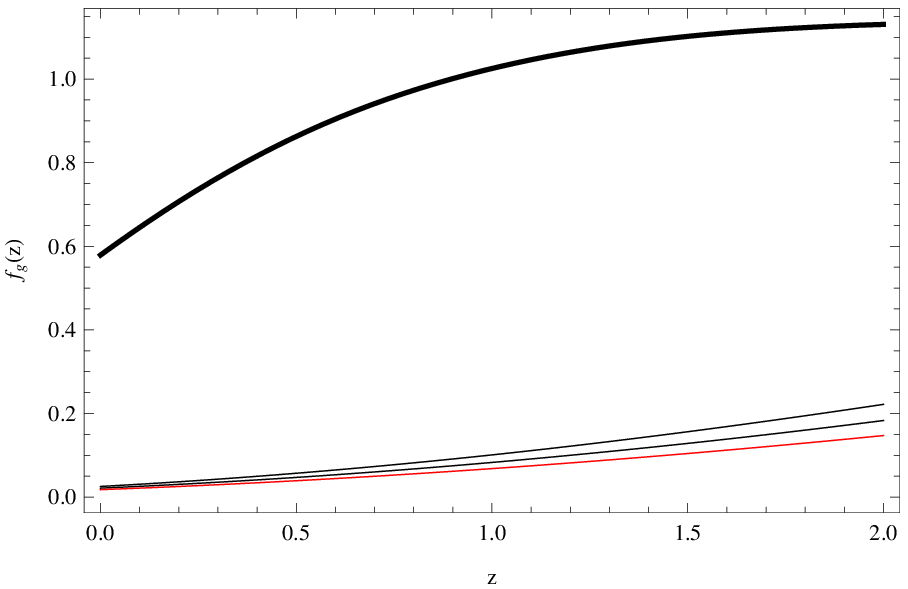}
\caption{Comparison of the growth factor $f_g(z)$ over z, for $k=0.1$Mpc$^{-1}$ for the model $\tanh (1+z)$ (black curve), for the model $\frac{2 z+5}{z+100}$ (red curve) and the model $\sqrt{2z+5}$(blue curve). In the left plot, the growth factors corresponding to the three models appear only while in the right plot the growth factor of the three models are compared to the standard non-mimetic $F(R)$ gravity model, for the exponential  with power-law term model (black thick curve).} \label{plot6}
\end{figure}
We now perform a numerical analysis of the differential equation (\ref{presenceofcoll}), using the initial condition $f_g(z_{fin}, k)=1$, with $z_{fin}=10$. Note that the differential equation (\ref{presenceofcoll}) is valid for specific values of $k$, in order that the sub-horizon approximation holds true.
\begin{table*}[h]
\small
\caption{\label{tableiiiv} Allowed values of the comoving wavenumber $k$ for the mimetic $F(R)$ models and for the ordinary $F(R)$ models.}
\begin{tabular}{@{}crrrrrrrrrrr@{}}
\tableline
\tableline
\tableline
Model & Allowed value of $k$
\\\tableline
Ordinary $F(R)$ Model &  $k>0.000115629$
 \\\tableline
$\mathcal{Q}(z)=\sqrt{2 z+5}$  &  $k>0.000133543$
\\\tableline
$\mathcal{Q}(z)=\frac{2 z+5}{z+100}$  &  $k>0.0000947621$ \\
\tableline
$\mathcal{Q}(z)=\tanh (z+1)$  &  $k>0.000110926$
\\\tableline
\tableline
 \end{tabular}
\end{table*}
In Table \ref{tableiiiv} we present the various values of the comoving wavenumber for the ordinary $F(R)$ gravity model and also for the three mimetic models, and we use allowed values of $k$ for the numerical analysis. Particularly, the value $k=0.1$Mpc$^{-1}$ is allowed for all models, so the results of the numerical analysis can be found in Fig. \ref{plot6}. As it can be seen in the right plot, all the mimetic models produce a growth factor which is quite well below the ordinary $F(R)$ gravity case (black thick curve). Also the three different models generate a different evolution of the growth factor, as it can be seen in the left plot. Hence, the growth factor can be a consistent criterion which differentiates the various models we studied in this paper.

\subsection{Alternative Scenarios and Further Perspectives}

\begin{figure}[h]
\centering
\includegraphics[width=15pc]{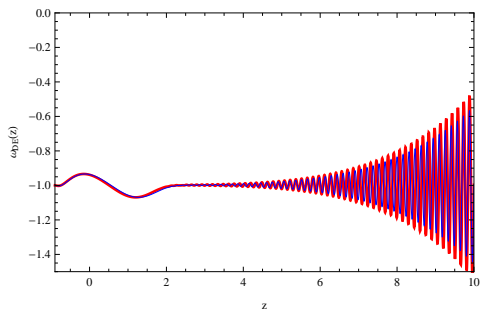}
\caption{Comparison of the dark energy equation of state $\Omega_{DE}(z)$, for the simple mimetic case $C_{\phi}/a^{3}$ (blue curve). The resulting dark energy oscillations are rescaled, depending on the choice of the parameter $C_{\phi}$. In this case we chose $C_{\phi}/\rho_{m}=10$. } \label{thorn}
\end{figure}

One case that we did not address in the previous sections is the case of mimetic $F(R)$ gravity without potential and Lagrange multiplier. This was studied in \cite{NO2}, and the resulting equations of motion for the FRW metric (\ref{frw}) are similar to Eqs. (\ref{enm1}), and (\ref{enm2}), with the difference that the potential and the Lagrange multiplier are replaced by a term $C_{\phi}/a^{3}$. It is exactly this term that mimics the dark matter, since it originates from the hidden conformal degree of freedom of the metric tensor. We performed an analysis for the cosmological evolution of the dark energy oscillations for this case, and the resulting picture is very similar to the ordinary $F(R)$ case, since the difference is just the presence of the free parameter $C_{\phi}$. So practically the only difference is that all the quantities are rescaled, for example the behavior of the dark energy parameter is presented in Fig. \ref{thorn}.
Finally, let us note that the presence of the mimetic potential and the Lagrange multiplier in the Lagrangian offers many possibilities for realization of various cosmologies. It is known that cosmographic studies indicate that the cosmological standard model behaves as follows,
 \cite{caponew},
\begin{equation}\label{capowner}
H(z)\sim \sqrt{\Omega_m(1+z)^3+\ln (\alpha+\beta z)}\, .
\end{equation}
In the mimetic model at hand, the Hubble rate in terms of the function $\mathcal{Q}(z)$ is given in Eq. (\ref{hubblepar}), so by choosing  the function $\mathcal{Q}(z)$ as follows,
\begin{equation}\label{hubbleparcaponew}
\mathcal{Q}(z)=\Omega_m(1+z)^3+\ln (\alpha+\beta z)-\tilde{m}^2(y_H(z)-\chi (z+1)^{4})\, ,
\end{equation}
then, the corresponding evolution is identical to (\ref{capowner}). Therefore, the mimetic case offers much freedom for model building.

\section*{Concluding Remarks}

In this paper we demonstrated that in the context of mimetic $F(R)$ gravity with Lagrange multiplier and mimetic potential, it is possible to solve in a kind of elegant way the problem of dark energy oscillations at late times. Particularly, in the standard $F(R)$ gravity approach, the dark energy equation of state oscillates strongly after $z\sim 3$, which is near the end of the matter domination era, and until $z\sim 0$, which is the present time. In the standard approach, the $F(R)$ gravity has to be modified by adding power-law modifications, however the problem is not completely solved. In the mimetic case, no power-law modifications are needed, and moreover by appropriately choosing the potential and the Lagrange multiplier, it is possible to minimize the amplitude of the oscillations. We performed a numerical analysis which showed in a clear way that the oscillations are strongly damped, so the mimetic potential and Lagrange multiplier solve this serious issue. In addition, by calculating the present day values of the mimetic dark energy density parameter $\Omega_{DE} (0)$, and the total effective equation of state parameter $\omega_{eff}(0)$, we demonstrated that the values are very close to the observational data and in some cases, full concordance with observations is achieved. Finally, we calculated the growth factor for all the mimetic models  and we compared its behavior as a function of the redshift $z$ with the ordinary $F(R)$ gravity case. As we showed, the various mimetic models have strong differences with the ordinary $F(R)$ and also there are differences between the growth factor of each model.

The mimetic $F(R)$ gravity framework proves to be very useful, since many cosmological evolutions can be realized in the context of this theory. The question is, can this theory be considered as a viable modified gravity candidate? Indeed it seems that with this theory, everything can be realized and someone could say that it lacks of predictability. From our point of view, mimetic gravity is more than a simple mathematical construction. It is appealing and in some sense ``economical'' since nothing new is added to the theory and the considerations involve the internal conformal degree of freedom of the metric. Moreover, compatibility with the observational data can be achieved and in addition, with the present work  we showed that the theoretical problem of dark energy oscillations is solved without adding by hand curvature correction terms. More importantly, the observational values of the dark energy density parameter and of the total effective equation of state parameter are in agreement or very close to the observational values of these parameters. But in all cases we have better agreement in comparison to the values corresponding to the ordinary $F(R)$ case, so the mimetic theory is subtly more appealing. Also the predicted growth factor of the mimetic models is quite lower in magnitude in comparison to the ordinary $F(R)$ case. Hence, by taking into account all the aforementioned features and also the ones appearing in the literature, one may find it difficult to claim that this theory is rather a mathematical construction, and the theory deserves more work in order to reveal all its fundamental features.

An interesting quite novel research stream could start by considering mimetic modifications of torsion-based theories of gravity, such as teleparallel theories of the form $f(T)$, with $T$ being the torsion scalar, see for example the recent work \cite{ref3}. This study would be strongly motivated by the fact that in the particular case $f(T)=T$, one gets the teleparallel equivalent of general relativity, in which case the equations of motion of the torsion and of the Einstein-Hilbert gravity coincide. However, in mimetic modifications of torsion theories, one should carefully reveal the conformal degree of freedom of the metric, by using the vierbeins appropriately. If this is true, a new theory with geometric originating dark matter would be obtained, so we defer this issue to a future work.

\section*{Acknowledgments}

This work is supported by MINECO (Spain), project
 FIS2013-44881 (S.D.O) and by Min. of Education and Science of Russia (S.D.O
and V.K.O).

\end{document}